\documentclass[prd,aps,showpacs,nofootinbib]{revtex4}%
\usepackage{amssymb}
\usepackage{graphicx}
\usepackage{amsmath}
\usepackage{amsfonts}%
\newcommand{\be}{\begin{equation}}
\newcommand{\ee}{\end{equation}}
\newcommand{\bq}{\begin{eqnarray}}
\newcommand{\eq}{\end{eqnarray}}
\begin{document}
\title{Classical Solutions in a Lorentz-violating Scenario of
Maxwell-Chern-Simons-Proca Electrodynamics}
\author{H. Belich Jr.$^{a,e}$}
\affiliation{$^{a}${\small {Universidade Federal do Esp\'{\i}rito Santo (UFES),
Departamento de F\'{\i}sica e Qu\'{\i}mica, Av. Fernando Ferrari, S/N
Goiabeiras, Vit\'{o}ria - ES, 29060-900 - Brasil}}}
\author{T. Costa-Soares$^{b,d}$}
\affiliation{$^{b}${\small {Universidade Federal de Juiz de Fora (UFJF), Col\'{e}gio
T\'{e}cnico Universit\'{a}rio, av. Bernardo Mascarenhas, 1283, Bairro
F\'{a}brica - Juiz de Fora - MG, 36080-001 - Brasil}}}
\author{M.M. Ferreira Jr.$^{c}$\ }
\affiliation{$^{c}${\small {Universidade Federal do Maranh\~{a}o (UFMA), Departamento de
F\'{\i}sica, Campus Universit\'{a}rio do Bacanga, S\~{a}o Luiz - MA, 65085-580
- Brasil}}}
\author{J.A. Helay\"{e}l-Neto$^{d}$}
\affiliation{$^{d}${\small {CBPF - Centro Brasileiro de Pesquisas F\'{\i}sicas, Rua Xavier
Sigaud, 150, CEP 22290-180, Rio de Janeiro, RJ, Brasil}}}
\author{}
\email{belich@cce.ufes.br, tcsoares@cbpf.br, manojr@ufma.br, helayel@cbpf.br. }

\begin{abstract}
Taking as starting point the planar model that arises from the dimensional
reduction of the Abelian-Higgs Carroll-Field-Jackiw model, we write down and
study the extended Maxwell equations and the associated wave equations for the
potentials. The solutions for these equations correspond to the usual ones for
the MCS-Proca system, supplemented with background-dependent correction terms.
In the case of a purely timelike background, exact algebraic solutions are
presented which possess a similar behavior to the MCS-Proca counterparts near
and far from the origin.\ On the other hand, for a purely spacelike
background, only approximate solutions are feasible. They consist of
non-trivial analytic expressions with a manifest evidence of spatial
anisotropy, which is consistent with the existence of a privileged direction
in space. These solutions also behave similarly to the MCS-Proca ones near and
far from the origin.

\end{abstract}
\maketitle

\section{ \ Introduction\-}

Lorentz- and CPT-violating theories in $\left(  1+3\right)  $-dimensions have
been object of intensive investigation in the latest years \cite{Jackiw}%
-\cite{Reduction2}. An odd-CPT Lorentz-violating model (with a
Chern-Simons-like term) was pioneering considered in the context of classical
electrodynamics by Carroll-Field-Jackiw \cite{Jackiw}, by setting up a simple
way to realize the CPT- and Lorentz-breakings in the framework of the Maxwell
theory. In a general perspective, an extension of the minimal $SU(3)\times
SU(2)\times U(1)$ standard model incorporating odd- and even-CTP terms was
developed by Colladay\ \& Kostelecky \cite{Colladay} as a low-energy limit of
a Lorentz covariant model valid at the Planck scale. This master model
undergoes a spontaneous symmetry breaking, generating an effective action that
incorporates Lorentz violation and keeps unaffected the $SU(3)\times
SU(2)\times U(1)$ gauge structure and the energy-momentum conservation. This
standard model extension (SME)\ has then been investigated under diverse
aspects \cite{Coleman}, \cite{Adam}.

The Carroll-Field-Jackiw model\cite{Jackiw}, in spite of predicting several
interesting new properties and a potentially rich phenomenology, is a model
plagued with some serious problems, like the absence of stability and
causality in the case of a purely timelike background, $v^{\mu}=($v$_{0},0). $
Even so, this theory has been fairly-well discussed under a number of
different aspects, like the following ones: (i) the birefringence (optical
activity of the vacuum), induced by the fixed background \cite{Jackiw}%
,\cite{Astrophys}, (ii) the investigation of radiative corrections
\cite{Chung}, (iii) the consideration of spontaneous breaking of U(1)-symmetry
in this framework \cite{Belich1}, (iv) the search for a supersymmetric
Lorentz-violating extension model \cite{Belich2}, (v) the study of vacuum
Cerenkov radiation \cite{Lehnert}, photon decay process \cite{Adam2}, and some
other points.

The great interest aroused by such an issue has motivated the study of
Lorentz-violating theories in lower dimensions. In this sense, a dimensional
reduction (to $D=1+2)$ of the Lorentz-breaking Maxwell Electrodynamics,
endowed with the Carroll-Field-Jackiw term ($\epsilon^{\mu\nu\kappa\lambda
}v_{\mu}A_{\nu}F_{\kappa\lambda}$) \cite{Jackiw}, has been recently performed
\cite{Reduction1}, yielding in a gauge invariant Planar Quantum
Electrodynamics (QED$_{3}$) composed by a Maxwell-Chern-Simons gauge sector, a
Klein-Gordon massless scalar field $\left(  \varphi\right)  $, and the fixed
3-vector $\left(  v^{\mu}\right)  $, responsible for the Lorentz-violation. As
for the physical consistency of this model, some of its general features have
been investigated. One has then verified that the complete model is stable and
preserves causality and unitarity without any restrictions \cite{Reduction1}.
Therefore, the full model supports a consistent quantization for both time-
and spacelike backgrounds. Furthermore, the classical motion equations and
solutions of this Lorentz-violating planar model were considered as well
\cite{Classical}, revealing interesting deviations in relation to the pure MCS
case, like the absence of screening in the electric sector for a purely
timelike background and manifest anisotropy for a purely spacelike background.

The Carroll-Field-Jackiw model has been also considered in the context of a
$U(1)$ spontaneous symmetry breaking, yielding an Abelian-Higgs
Lorentz-violating model in (1+3) dimensions endowed with stable vortex
configurations \cite{Belich1}. In a further work \cite{Reduction2}, one has
carried out the dimensional reduction of this master model to (1+2)
dimensions, obtaining a planar Lorentz-violating Lagrangian with the Higgs
sector. The consistency of this model was properly analyzed at the classical
level, revealing preservation of causality, stability and unitarity for both
time- and spacelike backgrounds, in a similar way to the case addressed to in
Ref. \cite{Reduction1}. Recently, one has also investigated the presence of
vortex configurations in this planar framework \cite{Vortex}, and it has been
found out that there may appear stable configurations of electrically charged
vortices which induce an Aharonov-Casher phase for \ neutral particles.

To study Lorentz-violating theories, we have adopted a general procedure that
consists in investigating and setting up its classical aspects before
addressing to the second-quantized case. With this program in mind, we have
discussed the consistency (causality, unitarity and stability) of the
Higgs-Carroll-Field-Jackiw model both in (1+3) and (1+2) dimensions, based on
the dispersion relations that are read off as poles of the propagators. This
task may indeed indicate the eventual presence of non-physical modes, such as
spacelike poles (tachyons) and negative-norm 1-particle states (ghosts). Once
one has thereby fixed the parameters of the model and selected the situations
for which the spectrum does not display unphysical excitations, then it is
sensible to carry out the second quantization of the system.

In the present paper, one follows this general procedure, now focusing the
attention on the Classical Electrodynamics that stems from the U(1) broken
phase of the planar version of the Abelian Higgs CFJ model. The main goal is
to describe the influence of the Lorentz-violating background on the solutions
associated with a system of point-like charges, bearing in mind the results
obtained in Ref. \cite{Classical}, which revealed the possibility of having
new physics induced by the presence of the background (as the vanishing of the
screening associated with the MCS Electrodynamics, for instance). In this
sense, one first writes down the tree-level Lagrangian of the U(1) broken
phase of the Higgs-Abelian model worked out in Ref. \cite{Reduction2}; it is
composed of a MCS-Proca gauge sector coupled to a Klein-Gordon massive field
by means of the Lorentz-violating term. The associated classical equations of
motion (the extended Maxwell equations) and wave equations (for the potential
$A^{\mu}$) are written in the sequel. Such equations correspond to the ones of
the usual MCS-Proca Electrodynamics supplemented by terms that depend on the
background vector. So, it might be expected that the solutions we find
correspond to the MCS-Proca ones corrected by background-dependent terms.
Indeed, this is the case. Proceeding further, solutions for field strengths
and potentials have been found for point-like charges (both for purely
timelike and spacelike backgrounds), exhibiting $v^{\mu}$-dependent
corrections with respect to the pure MCS-Proca counterparts.

Specifically, in the case of a purely timelike background, exact algebraic
solutions are attained by means of Fourier integrations. Both the scalar and
vector potentials are given in terms of linear combinations of modified Bessel
functions $(K_{0},K_{1})$ and behave at the origin and far from it in much the
same way as the pure MCS-Proca solutions; the difference always occurs at some
intermediary radial region. Since these Bessel functions decay exponentially,
it is evident that the associated solutions present a strong screening,
typical to the case where the physical intermediation is played only by
massive particles. It has also been noticed that the scalar potential $\left(
A_{0}\right)  $ exhibits a familiar form, similar to the MCS-Proca solution.
However, it may significantly differ from the latter potential in the case of
a small Proca mass $\left(  M_{A}/s<<1\right)  $ or a large background
$\left(  \text{v}_{0}\lesssim s\right)  ,$ in which case it becomes attractive
in some radial range. As for the vector potential $\left(  \mathbf{A}\right)
,$one is able to write down a solution rather similar to the MCS-Proca
counterpart, without qualitative alterations. However, these solutions may
differ substantially at intermediary distances for the case in which $\left(
\text{v}_{0}\lesssim s\right)  $. Plots are introduced to illustrate the
points alluded to here.

On \ the other hand, in the case of a purely spacelike background, the Fourier
integrations result to be no more exactly soluble, implying the necessity of
employing approximations which lead to algebraic solutions of great complexity
(in leading order in v$^{2}/s^{2}$). The presence of spatial-anisotropy
becomes a manifest property, in the form of correction terms with a clear
dependence on the angle determined by the fixed background ($\overrightarrow
{\text{v}}$).\ The scalar potential worked out consists of a complex
combination of Bessel and radial functions $(K_{0},rK_{1},K_{1}/r)$; its forms
near and far from the origin are qualitatively similar to the MCS-Proca case:
it vanishes for $r\rightarrow\infty$ and goes as $\ln r$ for $r\rightarrow0$.
Concerning the vector potential, it also appears as a lengthy combination of
Bessel and radial functions $(rK_{0},K_{0}/r,K_{1},K_{1}/r^{2}),$ exhibiting
anisotropy terms. In spite of the involved complexity, this potential presents
an identical behavior to the MCS-Proca counterpart near and away from the
origin. A Graphical analysis reveal that the presence of the background does
not amount to qualitative or sensitive modifications on the the MCS-Proca
solutions, since the small magnitude of the background compared with the
Chern-Simons parameters (v$^{2}/s^{2}<<1)$.

The method of investigation adopted here has yielded solutions for the
Klein-Gordon field as well, revealing an analogous structure to the scalar
potential both in the purely timelike and spacelike backgrounds. Moreover, it
is important to point out that such solutions recover the pure MCS-Proca
results in limit of a vanishing background $\left(  \text{v}^{\mu}=0\right)  ,
$ which is a necessary condition to attest the validity of the solutions found out.

In short, this paper is outlined as follows. In Sec. II, we present the basic
features of the reduced model, previously developed in ref. \cite{Reduction2}.
In Sec. III, the equations of motion, from which one derives the wave
equations for potentials and field strengths are presented. In Sec. IV, we
solve the equations for the scalar potential (in the static limit)\ for the
time- and space-like cases and discuss the results. In Sec. V, we solve the
differential equations for the vector potential according to the procedure
adopted in Sec. IV. In Sec. VI, we conclude by presenting our Final Remarks.

\section{The Dimensionally Reduced Lorentz-violating Model}

We take as starting point the Carroll-Field-Jackiw Lorentz-violating
electrodynamics minimally coupled to a scalar field sector, endowed with
spontaneous symmetry breaking \cite{Belich1}\footnote{Here one has adopted the
following metric conventions: $g_{\mu\nu}=(+,-,-,-)$ in $D=1+3,$ and
$g_{\mu\nu}=(+,-,-)$ in $D=1+2$.}:%

\begin{equation}
\mathcal{L}_{1+3}=-\frac{1}{4}F_{\hat{\mu}\hat{\nu}}F^{\hat{\mu}\hat{\nu}%
}+\frac{1}{4}\varepsilon^{\hat{\mu}\hat{\nu}\hat{\kappa}\hat{\lambda}}%
v_{\hat{\mu}}A_{\hat{\nu}}F_{\hat{\kappa}\hat{\lambda}}+(D^{\hat{\mu}}%
\phi)^{\ast}D_{\hat{\mu}}\phi-V(\phi^{\ast}\phi)-A_{\hat{\nu}}J^{\hat{\nu}},
\label{action1}%
\end{equation}
where $v^{\hat{\mu}}$ stands for the fixed background (associated with the
Lorentz-violation at the level of the particle frame) \cite{Kostelec1} and the
greek letters with hat, $\hat{\mu},$\ run from $0$\ to $3.$ Here, $D_{\hat
{\mu}}=(\partial_{\hat{\mu}}+ieA_{\hat{\mu}})$\ is the covariant derivative
which sets up the minimal coupling with the scalar field while $V(\phi^{\ast
}\phi)=m^{2}\phi^{\ast}\phi+\lambda(\phi^{\ast}\phi)^{2}$\ represents the
scalar potential responsible for spontaneous symmetry breaking. The
theoretical model of Lagrangian \ref{action1} was analyzed in Ref.
\cite{Belich1}, in which it has been shown that it is consistent (endowed with
causality and unitarity) only for a purely timelike background. \ 

We should now consider the dimensionally reduced version of this model, which
has been developed and analyzed in Ref. \cite{Reduction2}, where one can find
the motivations to consider it and details of the reduction process are
given.\textbf{\ }Applying the prescription of the dimensional reduction,
described in Refs. \cite{Reduction1},\cite{Reduction2}, on the Eq.
(\ref{action1}), one obtains the reduced planar Lagrangian: \ \ \ \ \ \
\begin{align}
\mathcal{L}_{1+2}  &  =-\frac{1}{4}F_{\mu\nu}F^{\mu\nu}+\frac{1}{2}%
\partial_{\mu}\varphi\partial^{\mu}\varphi+\frac{s}{2}\epsilon_{\mu\nu
k}A^{\mu}\partial^{\nu}A^{k}-\varphi\epsilon_{\mu\nu k}v^{\mu}\partial^{\nu
}A^{k}+(D_{\mu}\phi)^{\ast}(D_{\mu}\phi)-e^{2}\varphi^{2}(\phi^{\ast}%
\phi)\nonumber\\
&  -V\left(  \phi^{\ast}\phi\right)  -A_{\mu}J^{\mu}-\varphi J,
\label{Lagrange2}%
\end{align}
where the greek letters (now without hat) run from 0 to 2. The scalar field,
$\varphi,$\ is the remanent of the compactified coordinate of the vector
potential ($A^{(3)}=\varphi)$, here acting as a massless Klein-Gordon field.
The mixing Chern-Simons-like term,\ $\varphi\epsilon_{\mu\nu k}v^{\mu}%
\partial^{\nu}A^{k}$, \ in spite of covariant in form, is not
Lorentz-invariant in the particle-frame, in which the fixed ($v^{\mu}%
)$\ background does not boost as a 3-vector. The Lagrangian \ref{Lagrange2}
represents a field model endowed with Lorentz violation and spontaneous
symmetry breaking, which may constitute a theoretical framework useful to
analyze planar vortex configurations. Its components present the following
mass dimension: $[A^{\mu}]=[\varphi]=1/2,$\ $[s]=[v^{\mu}]=1,[J^{\mu}]=5/2.$\ 

\ Having established the planar Lorentz-violating model, we can now consider
the spontaneous symmetry breaking process, which provide mass to the gauge and
scalar fields \cite{Reduction2}. Once we are bound to a tree-level analysis,
we then retain only the bilinear terms, so that the planar Lagrangian takes
the form:
\begin{equation}
\mathcal{L}_{1+2}^{broken}=-\frac{1}{4}F_{\mu\nu}F^{\mu\nu}+\frac{1}%
{2}\partial_{\mu}\varphi\partial^{\mu}\varphi-\frac{1}{2}M_{A}^{2}\varphi
^{2}+\frac{s}{2}\epsilon_{\mu\nu k}A^{\mu}\partial^{\nu}A^{k}-\varphi
\epsilon_{\mu\nu k}v^{\mu}\partial^{\nu}A^{k}+\frac{1}{2}M_{A}^{2}A_{\mu
}A^{\mu}-A_{\mu}J^{\mu}-\varphi J,\label{Lagrange3}%
\end{equation}
where: $M_{A}^{2}=2e^{2}\langle\phi\phi\rangle,$ with $\langle\phi\phi
\rangle\ $being the vacuum expectation value of the scalar field. The
tree-level Lagrangian above represents a theoretical model \ composed of a
Maxwell-Chern-Simons-Proca gauge sector, the massive Klein-Gordon field and
the Lorentz-violating mixing term. This is the Maxwell-Chern-Simons-Proca
electrodynamics corrected by the presence of the fixed background. The Higgs
field is not\ considered in the Lagrangian above once we work in the U(1)
broken phase and the unitary gauge has been chosen; as a consequence, this
field has its own kinetic term and does not mix, as far as only bilinear terms
are considered, with the gauge-field part of the action.

In Ref. \cite{Reduction2}, the field propagators related to Lagrangian
(\ref{Lagrange3}) were properly evaluated and taken as starting point to
analyze the consistency of this planar model, which has revealed to be totally
causal and unitary for both timelike and spacelike backgrounds. Indeed, no
problem concerning causality and unitarity was found out.

\section{ Wave Equations for Potentials and Field Strengths}

We now go on writing the extended wave equations which govern the behavior for
the potentials components and field strengths of the scalar electrodynamics
stated in Lagrangian (\ref{Lagrange3}), from which there follows two
Euler-Lagrangian equations of motion:%

\begin{align}
\partial_{\nu}F^{\mu\nu}  &  =s\varepsilon^{\mu\nu\rho}\partial_{\nu}A_{\rho
}+\varepsilon^{\mu\nu\rho}v_{\nu}\partial_{\rho}\varphi+M_{A}^{2}A^{\mu
}-J^{\mu},\label{motion1}\\
(\square+M_{A}^{2})\varphi &  =-\epsilon_{\mu\nu k}v^{\mu}\partial^{\nu}%
A^{k}-J, \label{motion2}%
\end{align}
which lead to the extended Maxwell equations:
\begin{align}
\overrightarrow{\nabla}\times\overrightarrow{E}+\partial_{t}B\text{ }  &
=0,\label{Maxwell1}\\
\partial_{t}\overrightarrow{E}-\nabla^{\ast}B\text{ \ }  &  =-\overrightarrow
{j}-s\overrightarrow{E}^{\ast}-\left(  \overrightarrow{\text{v}}^{\ast
}\partial_{t}\varphi+\text{v}_{0}\overrightarrow{\nabla}^{\ast}\varphi\right)
+M_{A}^{2}\overrightarrow{A},\label{Maxwell2}\\
\overrightarrow{\nabla}\cdot\overrightarrow{E}\text{ }-\text{ }sB\text{ }  &
=\rho-M_{A}^{2}A_{0}+\overrightarrow{\text{v}}\times\overrightarrow{\nabla
}\varphi,\label{Maxwell3}\\
(\square+M_{A}^{2})\varphi &  =\text{v}_{0}\overrightarrow{\nabla}%
\times\overrightarrow{A}-\overrightarrow{\text{v}}\times\overrightarrow{E}-J.
\label{Maxwell4}%
\end{align}
The first of these equations\ is the non-covariant form of the Bianchi
identity $(\partial_{\mu}F^{\mu\ast}=0)$\footnote{In $D=1+2$ the dual tensor,
defined as $F^{\mu\ast}=\frac{1}{2}\epsilon^{\mu\nu\alpha}F_{\nu\alpha},$ is a
3-vector given by: $F^{\mu\ast}=(B,-\overrightarrow{E}^{\ast}).$ Here one
adoptes the following convection: $\epsilon_{012}=\epsilon^{012}=\epsilon
_{12}=\epsilon^{12}=1. $ The symbol $(^{\ast})$ designates the dual of a
vector; it transforms an ARBITRARY 2-vector $\mathbf{A}=(A_{x},A_{y})$ at the
form: $\mathbf{A}^{\ast}=(A_{y},-A_{x}).$}. Motion equation (\ref{motion1})
yields the two inhomogeneous ones, while Eq. (\ref{motion2}) leads to last
one. \ From such equations, one readily determines the mass dimension of the
field strengths, namely: $[E]=[B]=3/2.$ The original four-dimensional
Lorentz-breaking model is gauge invariant \cite{Jackiw}, property transferred
also for the planar model. It may be directly demonstrated from Eq.
(\ref{motion1}); one easily obtains: $\partial_{\mu}J^{\mu}=-\varepsilon
^{\mu\nu\rho}\partial_{\mu}v_{\nu}\partial_{\rho}\varphi$. Whenever $v^{\mu}$
is constant or has a null rotational $(\varepsilon^{\mu\nu\rho}\partial_{\mu
}v_{\nu}=0)$, this equation leads to the conventional current-conservation
law, $\partial_{\mu}J^{\mu}=0$, consistent with gauge invariance.

From pure algebraic manipulation of the Maxwell equations, one gets that the
fields $B$, $\overrightarrow{E,}$ satisfy second-order inhomogeneous wave
equations:
\begin{align}
(\square+s^{2}+M_{A}^{2})B  &  =-s\rho+\overrightarrow{\nabla}\times
\overrightarrow{j}+sM_{A}^{2}A_{0}-s\overrightarrow{\text{v}}\times
\nabla\varphi-\partial_{t}\left(  \nabla\varphi\right)  \times\overrightarrow
{\text{v}}^{\ast}-\text{v}_{0}\nabla^{2}\varphi,\label{B1}\\
(\square+s^{2}+M_{A}^{2})\overrightarrow{E}  &  =-\overrightarrow{\nabla}%
\rho-\partial_{t}\overrightarrow{j}+s\overrightarrow{j}^{\ast}-\overrightarrow
{\nabla}(\overrightarrow{v}\times\overrightarrow{\nabla\varphi}%
)-s\overrightarrow{v}\left(  \partial_{t}\varphi\right)  -s\text{v}%
_{0}\overrightarrow{\nabla}\varphi-sM_{A}^{2}\overrightarrow{A}^{\ast
}\nonumber\\
&  -\overrightarrow{\text{v}}^{\ast}\partial_{t}^{2}\varphi+\text{v}%
_{0}\overrightarrow{\nabla}^{\ast}\left(  \partial_{t}\varphi\right)  .
\label{E1}%
\end{align}
Similarly to the\ classical MCS model, the potential components $(A_{0}%
,\overrightarrow{A})$ obey fourth-order wave inhomogeneous equations:%

\begin{align}
\left[  \square(\square+s^{2}+2M_{A}^{2})+M_{A}^{2}\right]  A_{0}  &  =\left(
\square+M_{A}^{2}\right)  [\rho+(\overrightarrow{\text{v}}\times
\overrightarrow{\nabla\varphi})]+s\left(  \partial_{t}\overrightarrow
{\nabla\varphi}\right)  \times\overrightarrow{\text{v}}^{\ast}%
+s\overrightarrow{\nabla}\times\overrightarrow{j}-s\text{v}_{0}\nabla
^{2}\varphi,\label{Ascalar}\\
\left[  \square(\square+s^{2}+2M_{A}^{2})+M_{A}^{2}\right]  \overrightarrow
{A}  &  =\left(  \square+M_{A}^{2}\right)  (\overrightarrow{j}+\overrightarrow
{\text{v}}\partial_{t}\varphi+\text{v}_{0}\overrightarrow{\nabla}^{\ast
}\varphi)-s\overrightarrow{\nabla}^{\ast}\rho-s\partial_{t}\overrightarrow
{j}^{\ast}+s\overrightarrow{\text{v}}\left(  \partial_{t}^{2}\varphi\right)
\nonumber\\
&  -s\text{v}_{0}\overrightarrow{\nabla}\left(  \partial_{t}\varphi\right)
-s\left(  \overrightarrow{\nabla}(\overrightarrow{\text{v}}\times
\overrightarrow{\nabla\varphi})\right)  ^{\ast}, \label{Avector}%
\end{align}
It should be pointed out that the complexity of the inhomogeneous sector is
directly related to the presence of the background 3-vector in the Lagrangian
(\ref{Lagrange3}). In the absence of the background ($v^{\mu}\longrightarrow
0),$ it is useful to verify that wave equations (\ref{B1}, \ref{E1},
\ref{Ascalar}, \ref{Avector}) reduce to their classical MCS-Proca usual form:
\begin{align}
\left[  \square(\square+s^{2}+2M_{A}^{2})+M_{A}^{4}\right]  A_{0}  &  =\left(
\square+M_{A}^{2}\right)  \rho+s\overrightarrow{\nabla}\times\overrightarrow
{j},\label{MCSP1}\\
\left[  \square(\square+s^{2}+2M_{A}^{2})+M_{A}^{4}\right]  \overrightarrow
{A}  &  =\left(  \square+M_{A}^{2}\right)  \overrightarrow{j}-s\overrightarrow
{\nabla}^{\ast}\rho-s\partial_{t}\overrightarrow{j}^{\ast},\text{\ }%
\label{MCSP2}\\
\lbrack\square+s^{2}+M_{A}^{2}]\overrightarrow{E}  &  =-\overrightarrow
{\nabla}\rho-\partial_{t}\overrightarrow{j}+s\overrightarrow{j}^{\ast}%
-sM_{A}^{2}\overrightarrow{A}^{\ast},\label{MCSP3}\\
\lbrack\square+s^{2}+M_{A}^{2}]B  &  =-s\rho+\overrightarrow{\nabla}%
\times\overrightarrow{j}+sM_{A}^{2}A_{0}.
\end{align}
For a static point-like charge distribution, the wave equations above present
the following solutions:
\begin{align}
A_{0}(r)  &  =\left(  e/2\pi\right)  [c_{+}K_{0}(m_{+}r)+c_{-}K_{0}%
(m_{-}r)],\text{\ \ }\label{MCSP5}\\
\overrightarrow{A}(r)  &  =-\left(  e/2\pi\right)  c[m_{+}K_{1}(m_{+}%
r)-m_{-}K_{1}(m_{-}r)]\overset{\wedge}{r^{\ast}},\label{MCSP6}\\
\text{\ }\overrightarrow{E}  &  =-\left(  e/2\pi\right)  [c_{+}m_{+}%
K_{1}(m_{+}r)+c_{-}m_{-}K_{1}(m_{-}r)]\overset{\wedge}{r};\label{MCSP7}\\
B(r)  &  =-\left(  e/2\pi\right)  c[m_{+}^{2}K_{0}(m_{+}r)-m_{-}^{2}%
K_{0}(m_{-}r)], \label{MCSP8}%
\end{align}
with:
\begin{align}
c_{\pm}  &  =\frac{1}{2}\left[  1\pm\frac{s}{\sqrt{s^{2}+4M_{A}^{2}}}\right]
,\text{ }c=\frac{1}{\sqrt{s^{2}+4M_{A}^{2}}},\\
m_{\pm}^{2}  &  =\frac{1}{2}\left[  (s^{2}+2M_{A}^{2})\pm s\sqrt{s^{2}%
+4M_{A}^{2}}\right]  . \label{m}%
\end{align}
Near the origin these solutions behave as:\ $A_{0}(r)\rightarrow-\left(
e/2\pi\right)  \ln r,\overrightarrow{A}(r)\rightarrow0,\overrightarrow
{E}\rightarrow\left(  e/2\pi\right)  \overset{\wedge}{r}/r,B(r)\rightarrow
\left(  e/2\pi\right)  s\ln r.$ Far from the origin, all these solutions
vanish according to the asymptotic exponentially-decaying behavior of the
Bessel functions. \ The solutions above will be used as a reference to help
the identification of the contributions stemming from the presence of the
background to the MCS-Proca electrodynamics stated in Eq. (\ref{Lagrange3}).

\section{Solutions for the scalar potential and electric field in the static
limit}

In this section, we focus on the solutions for the scalar potential and
electric field for a static point-like charge for the case of both timelike
and spacelike Lorentz-violating backgrounds. These solutions are worked out
from the differential equations (\ref{Maxwell4}) and (\ref{Ascalar}), which
(in the static limit) become a coupled system of two differential equations.

\subsection{The external vector is purely time-like: $v^{\mu}=($v$_{0},0)$}

In the case of a static configuration, Eqs. (\ref{Maxwell4}),(\ref{Ascalar})
are reduced to the form:
\begin{align}
\lbrack\nabla^{2}(\nabla^{2}-s^{2}-2M_{A}^{2})+M_{A}^{4}]A_{0}+s\text{v}%
_{0}\nabla^{2}\varphi &  =-(\nabla^{2}-M_{A}^{2})\rho,\label{Ascalar2}\\
\text{v}_{0}(\nabla^{2}-M_{A}^{2})A_{0}-s(\nabla^{2}-M_{A}^{2})\varphi &
=-\text{v}_{0}\rho,
\end{align}
which consist of a system of two coupled linear differential equations. It is
possible to decouple these two equations to get the following ones:\
\begin{align}
\lbrack\nabla^{2}(\nabla^{2}-s^{2}-2M_{A}^{2})+M_{A}^{4}+\text{v}_{0}%
^{2}\nabla^{2}](\nabla^{2}-M_{A}^{2})A_{0}  &  =-\left[  (\nabla^{2}-M_{A}%
^{2})(\nabla^{2}-M_{A}^{2})+\text{v}_{0}^{2}\nabla^{2}\right]  \rho
,\label{Ascalar3}\\
s[\nabla^{2}(\nabla^{2}-s^{2}-2M_{A}^{2})+M_{A}^{4}+\text{v}_{0}^{2}\nabla
^{2}][\nabla^{2}-M_{A}^{2}]\varphi &  =-\text{v}_{0}\{[\nabla^{2}(\nabla
^{2}-s^{2}-2M_{A}^{2})+M_{A}^{4}]+\nonumber\\
&  +\text{v}(\nabla^{2}-M_{A}^{2})(\nabla^{2}-M_{A}^{2})\}\rho.
\end{align}

In order to solve Eq. (\ref{Ascalar3}), one proposes a point-like
charge-density distribution, $\rho\left(  r\right)  =e\delta(r),$ and a
Fourier-transform representation for the scalar potential, $A_{0}(r)=\frac
{1}{(2\pi)^{2}}\int d^{2}\overrightarrow{k}e^{i\overrightarrow{k}%
.\overrightarrow{r}}\widetilde{A}_{0}(k),$ so that there follows a
Bessel-K$_{0}$ solution:
\begin{equation}
A_{0}(r)=\frac{e}{(2\pi)}\left[  \left(  A_{+}+\text{v}_{0}^{2}B_{+}\right)
K_{0}\left(  M_{+}r\right)  +\left(  A_{-}+\text{v}_{0}^{2}B_{-}\right)
K_{0}\left(  M_{-}r\right)  -\text{v}_{0}^{2}\left(  B_{+}+B_{-}\right)
K_{0}\left(  M_{A}r\right)  \right]  , \label{Azero3}%
\end{equation}
where the involved constants are give below:
\begin{align}
A_{\pm}  &  =\frac{1}{2}\left[  1\pm\frac{(s^{2}-\text{v}_{0}^{2})}%
{\sqrt{(s^{2}-\text{v}_{0}^{2})(s^{2}-\text{v}_{0}^{2}+4M_{A}^{2})}}\right]
,\label{a1}\\
B_{\pm}  &  =\left[  \frac{2T_{\pm}}{(s^{2}-\text{v}_{0}^{2})\pm\sqrt
{(s^{2}-\text{v}_{0}^{2})(s^{2}-\text{v}_{0}^{2}+4M_{A}^{2})}}\right]  ,\text{
\ }\label{b1}\\
T_{\pm}  &  =\frac{1}{2}\left[  1\pm\frac{(s^{2}-\text{v}_{0}^{2}+2M_{A}^{2}%
)}{\sqrt{(s^{2}-\text{v}_{0}^{2})(s^{2}-\text{v}_{0}^{2}+4M_{A}^{2})}}\right]
,\label{t1}\\
M_{\pm}^{2}  &  =\frac{1}{2}\left[  (s^{2}-\text{v}_{0}^{2}+2M_{A}^{2}%
)\pm\sqrt{(s^{2}-\text{v}_{0}^{2})(s^{2}-\text{v}_{0}^{2}+4M_{A}^{2})}\right]
. \label{M1}%
\end{align}
The electric field, derived from Eq. (\ref{Azero3}), is given simply by:
\begin{equation}
\overrightarrow{E}(r)=-\frac{e}{(2\pi)}\left[  -\left(  A_{+}+\text{v}_{0}%
^{2}B_{+}\right)  M_{+}K_{1}\left(  M_{+}r\right)  -\left(  A_{-}+\text{v}%
_{0}^{2}B_{-}\right)  M_{-}K_{1}\left(  M_{-}r\right)  +\text{v}_{0}%
^{2}\left(  B_{+}+B_{-}\right)  M_{A}K_{1}\left(  M_{A}r\right)  \right]
\overset{\wedge}{r}. \label{E2}%
\end{equation}
Both the electric field and scalar potential expressions present nearly the
same functional behavior as the corresponding MCS-Proca, given by Eqs.
(\ref{MCSP5}), (\ref{MCSP7}), when v$_{0}/s<<1$ or $M_{A}/s\sim1,$ as it shall
be explained below. The presence of the background is not decisive to
determine qualitative modifications in their form both near and far from the
origin. Indeed, in the limit of short distances $\left(  r\ll1\right)  ,$ the
scalar potential (\ref{Azero3}) exhibits a purely logarithmic behavior,
whereas the electric field (\ref{E2}) goes as a $1/r$ function,
\begin{equation}
A_{0}(r)=-\left(  \frac{e}{2\pi}\right)  \ln r,\text{ \ \ \ \ }\overrightarrow
{E}(r)=-\left(  \frac{e}{2\pi}\right)  \frac{1}{r}\overset{\wedge}{r},
\end{equation}
which reveals the repulsive character of expression (\ref{Azero3}) near the
origin. It is interesting to remark that, in this limit, all the background
corrections drop out, and do not lead to modifications to the MCS-Proca
behavior near the origin. \ Far from the origin $\left(  r\rightarrow
\infty\right)  ,$ both the scalar potential and electric fields decay
exponentially, showing an entirely screened behavior.

In a general sense, the background only seems to promote a damping in the
screening of the solutions, increasing then their range. The smaller $are$ the
factors $M_{\pm},$ the larger is the range. As far as $M_{\pm}^{2}<m_{\pm}%
^{2},$ the range of these new solutions is larger than the MCS-Proca
correspondents. Despite the functional similarity between the potentials
(\ref{MCSP5}) and (\ref{Azero3}), they may differ substantially in two clear
situations: (i) the Proca mass is small in comparison with the other mass
parameters $\left(  s,M_{+},M_{-}\right)  $ of this solution, which turns the
term $K_{0}(M_{A}r)$ dominant and reverses the behavior of the scalar
potential; (ii) the modulus of v$_{0}$ is near the topological mass $\left(
\text{v}_{0}/s\lesssim1\right)  ,$ in which regime the influence of the
background upon the solutions is maximal. The graphs in Fig. 1 and Fig. 2
illustrate these cases:%

\begin{figure}
[h]
\begin{center}
\fbox{\includegraphics[
trim=0.000000in 0.000000in 0.317007in 0.324654in,
height=2.7103in,
width=3.7135in
]%
{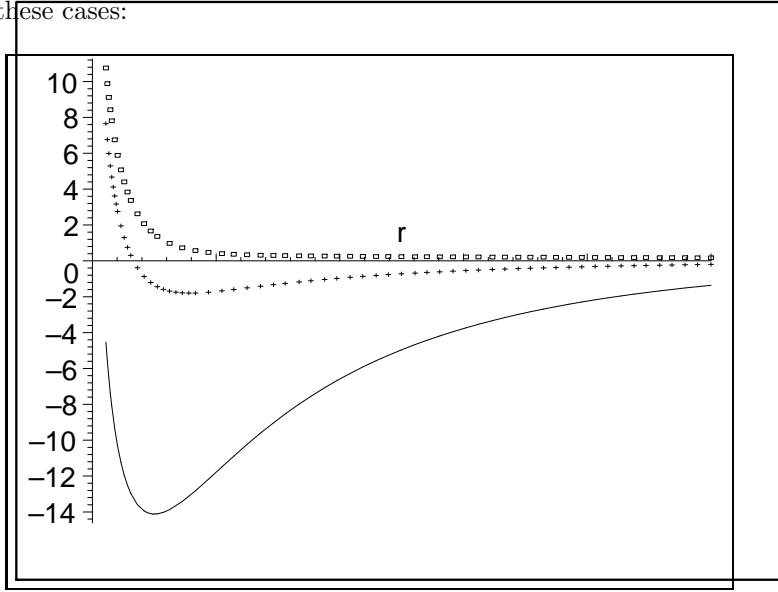}%
}\caption{Simultaneous plot of the pure MCS-Proca potential (box dotted line)
for $s=20,M_{A}=2;$ the scalar potential (circle dotted line) for
$s=20,M_{A}=2,$v$_{0}=8;$~the scalar potential for $s=20,M_{A}=2,$v$_{0}=15$
(continuos line).}%
\end{center}
\end{figure}

In Fig.1, one shows three curves for a small value of the Proca mass $\left(
M_{A}=2\right)  $. One then verifies that the closer v$_{0}$ is from the
$s-$value (in this case $s=20)$, the bigger the deviation from the pure
MCS-Proca behavior, as illustrated by the continuous curve. As the value of
v$_{0}$ decreases, the scalar potential tends to the MCS-Proca behavior, as
shown by the intermediate cross dotted line. For v$_{0}=0,$ we obviously
recover the pure MCS-Proca behavior, depicted by the box dotted line. The
scalar potential is negative in some radial extent due to the role played by
the term $-K_{0}(M_{A}r),$ which becomes dominant over the $K_{0}(M_{\pm}r)$
terms for $M_{A}<<M_{\pm}.$ Hence, the attractiveness here observed is
ascribed to the smallness of the ratio $M_{A}/s.$ In Fig. 2, the same kind of
simultaneous plot is displayed for a larger value of $M_{A},$ where from one
notes that the deviations from the MCS-Proca behavior are strongly attenuated
whenever the ratio $M_{A}/s$ increases.%

\begin{figure}
[h]
\begin{center}
\fbox{\includegraphics[
height=3.0364in,
width=4.0318in
]%
{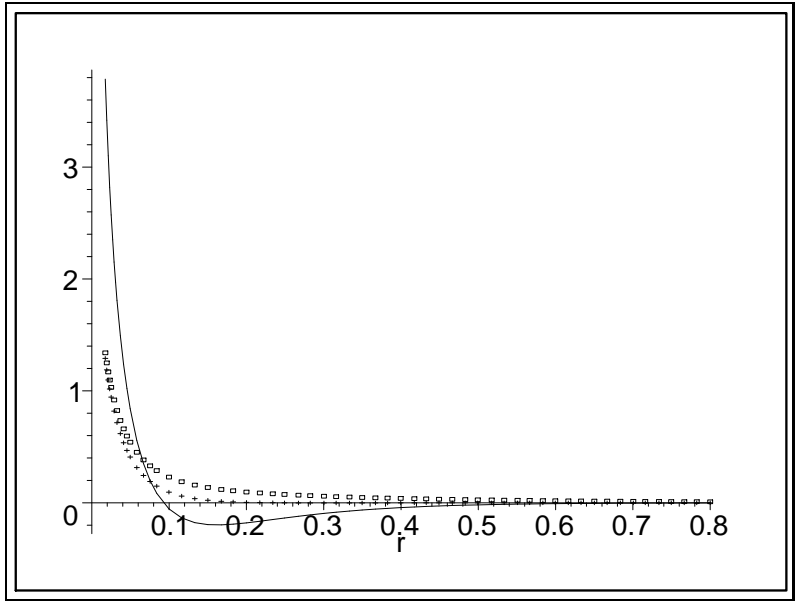}%
}\caption{Box dotted line: plot of the pure MCS-Proca potential for
$s=20,M_{A}=8;$ Cross dotted line: plot of the scalar potential for
$s=20,M_{A}=8,$v$_{0}=8;$~Continuos line: plot of the scalar potential for
$s=20,M_{A}=8,$v$_{0}=17.$}%
\end{center}
\end{figure}

Both Fig. 1 and Fig. 2 show that the scalar potential is always repulsive near
the origin and decays exponentially for large distances. Its behavior is very
similar to the MCS-Proca one in the cases in which v$_{0}/s<<1$ or
$M_{A}/s\sim1,$ but deviates substantially from it in the case one has
$M_{A}<<M_{\pm}$ or v$_{0}/s\lesssim1,$ for which one observes a potential
that becomes attractive at intermediary distances. \ As a final point, it is
important to remark that in the limit of a vanishing background (v$_{0}%
\longrightarrow0),$ one trivially recovers the MCS-Proca solutions, since
$A_{\pm}\rightarrow c_{\pm\text{ }}$in this situation.

\subsection{The external vector is\ purely\ space-like: $v^{\mu}=(0,
$\textbf{v}$)$}

In this case, one should consider Eqs. (\ref{Maxwell4}),(\ref{Ascalar}), which
in the static regime are written at the form:\
\begin{align}
\lbrack\nabla^{2}(\nabla^{2}-s^{2}-2M_{A}^{2})+M_{A}^{4}]A_{0}+\left(
\nabla^{2}-M_{A}^{2}\right)  (\overrightarrow{\text{v}}\times\overrightarrow
{\nabla\varphi})  &  =-(\nabla^{2}-M_{A}^{2})\rho\label{Azero6a}\\
(\overrightarrow{\text{v}}\times\overrightarrow{\nabla})A_{0}+(\nabla
^{2}-M_{A}^{2})\varphi &  =0,
\end{align}
Decoupling these equations, one attains:
\begin{align}
\lbrack\nabla^{2}(\nabla^{2}-s^{2}-2M_{A}^{2})+M_{A}^{4}-(\overrightarrow
{\text{v}}^{\ast}\cdot\overrightarrow{\nabla})(\overrightarrow{\text{v}}%
^{\ast}\cdot\overrightarrow{\nabla})]A_{0}  &  =-(\nabla^{2}-M_{A}^{2}%
)\rho,\label{Azero6}\\
\lbrack\nabla^{2}(\nabla^{2}-s^{2}-2M_{A}^{2})+M_{A}^{4}-(\overrightarrow
{\text{v}}^{\ast}\cdot\overrightarrow{\nabla})(\overrightarrow{\text{v}}%
^{\ast}\cdot\overrightarrow{\nabla})]\varphi &  =-(\overrightarrow{\text{v}%
}^{\ast}\cdot\overrightarrow{\nabla})\rho. \label{phi6}%
\end{align}

Starting from a point-like charge density distribution, $\rho\left(  r\right)
=e\delta(r),$ and proposing again a Fourier-transform representation for the
scalar potential, the solution will be given by the general the integral expression:%

\begin{equation}
A_{0}(r)=\frac{e}{\left(  2\pi\right)  ^{2}}\int_{0}^{\infty}\frac
{\mathbf{k}d\mathbf{k}}{\left[  \mathbf{k}^{2}+R_{+}^{2}\right]  }\int
_{0}^{2\pi}P_{+}e^{ikr\cos\varphi}d\varphi-\int_{0}^{\infty}\frac
{\mathbf{k}d\mathbf{k}}{\left[  \mathbf{k}^{2}+R_{-}^{2}\right]  }\int
_{0}^{2\pi}P_{-}e^{ikr\cos\varphi}d\varphi,
\end{equation}
where:
\begin{equation}
P_{\pm}=\frac{1}{2}\left[  1\pm\frac{(s^{2}+\text{v}^{2}\sin^{2}\alpha)}%
{\sqrt{(s^{2}+\text{v}^{2}\sin^{2}\alpha)(s^{2}+\text{v}^{2}\sin^{2}%
\alpha+4M_{A}^{2})}}\right]  , \label{c}%
\end{equation}%
\begin{equation}
R_{\pm}^{2}=\frac{1}{2}\left[  (s^{2}+2M_{A}^{2}+\text{v}^{2}\sin^{2}%
\alpha)\pm\sqrt{(s^{2}+\text{v}^{2}\sin^{2}\alpha)(s^{2}+4M_{A}^{2}%
+\text{v}^{2}\sin^{2}\alpha)}\right]  , \label{R}%
\end{equation}
and $\alpha$ is the angle defined by the relation $\cos\alpha=\overrightarrow
{\text{v}}\cdot\overrightarrow{k}/$v$k,$ that is, the angle determined by the
external background ($\overrightarrow{\text{v}})$ and transfer momentum
$(\overrightarrow{k})$. The fact the constants $\ P_{\pm},R_{\pm}$ depend on
the angle variable $\left(  \alpha\right)  $ implies that the integrations
above can not be exactly solved. An exact result was not found for these
integrations, but a sensible approximation can be performed in order to solve
them algebraically. Indeed, considering the condition $s^{2}>>$v$^{2},$ some
approximations are necessary so that the integration indicated becomes
feasible. In this regime, one has:%

\begin{align}
P_{\pm}  &  \simeq\frac{1}{2}[1\pm s/\gamma\pm\left(  2M_{A}^{2}\text{v}%
^{2}/s\gamma^{3}\right)  \sin^{2}\alpha],\label{App4}\\
\frac{1}{\left[  \mathbf{k}^{2}+R_{\pm}^{2}\right]  }  &  \simeq\frac
{1}{\left[  \mathbf{k}^{2}+m_{\pm}^{2}\right]  }\mp\frac{m_{\pm}^{2}}{s\gamma
}\frac{\text{v}^{2}\sin^{2}\alpha}{\left[  \mathbf{k}^{2}+m_{\pm}^{2}\right]
^{2}}, \label{App5}%
\end{align}
with: $m_{\pm}^{2}=\left[  s^{2}+2M_{A}^{2}\pm s\gamma\right]  /2,$
and$\ \gamma=\sqrt{s^{2}+4M_{A}^{2}}.$ It should be remarked that the factors
$m_{\pm}^{2}$ are exactly the ones that appear in the MCS-Proca solutions,
given by Eq. (\ref{m}). Here, one considers as well the angle between
$\overrightarrow{\text{v}}$ and $\overrightarrow{r}$, given by: $\cos\beta=$
$\overrightarrow{\text{v}}\cdot\overrightarrow{r}/$v$r,$ where $\beta=cte.$
\ While the background vector,\ $\overrightarrow{\text{v}},$ sets up a fixed
direction in space, the coordinate vector, $\overrightarrow{r},$ defines the
position where the potentials are to be measured; so, $\beta$ is the (fixed)
angle that indicates the directional dependence of the fields in relation to
the background direction. Being confined to the plane, these angles satisfy a
simple relation: $\alpha=\varphi-\beta,$ which allows the evaluation of the
angular integration on the $\varphi-$variable, based on the following
expression: $\sin^{2}\alpha=\cos^{2}\beta-(\cos2\beta)\cos^{2}\varphi
+c_{3}\sin2\varphi$. Considering all that, one has:
\begin{align}
\int_{0}^{2\pi}\left[  P_{-}e^{ikr\cos\varphi}\right]  d\varphi &  \simeq
\frac{\left(  2\pi\right)  }{2}\left[  \left(  1-s/\gamma-\epsilon\sin
^{2}\beta\right)  J_{0}(\mathbf{k}r)+\epsilon\cos2\beta J_{1}(\mathbf{k}%
r)/\left(  \mathbf{k}r\right)  \right]  ,\\
\int_{0}^{2\pi}\left[  P_{+}e^{ikr\cos\varphi}\right]  d\varphi &  \simeq
\frac{\left(  2\pi\right)  }{2}\left[  \left(  1+s/\gamma+\epsilon\sin
^{2}\beta\right)  J_{0}(\mathbf{k}r)-\epsilon\cos2\beta J_{1}(\mathbf{k}%
r)/\left(  \mathbf{k}r\right)  \right]  ,
\end{align}
with: $\epsilon=2M_{A}^{2}$v$^{2}/s\gamma^{3}.$ However, the task is not
complete yet.\ In order to solve the integration in dk, it is essential to
notice that the terms $R_{\pm}^{2}$, given in Eq. (\ref{R}), are also
dependent on the angle variable, requiring the use of another suitable
approximation, given in Eq. (\ref{App5}). Taking into account the above
angular integrations, one then carries out the k-integrations, arriving (at
first order in v$^{2}/s^{2}$) a lengthy expression for the scalar potential,
namely: \
\begin{equation}
A_{0}(r)=\frac{e}{2(2\pi)}\biggl\{\delta_{+}K_{0}(m_{+}r)+\delta_{-}%
K_{0}(m_{-}r)-\sigma_{+}(r)K_{1}(m_{+}r)+\sigma_{-}(r)K_{1}(m_{-}r)\biggr\},
\label{Azero4}%
\end{equation}
where:
\[
\delta_{\pm}=\left[  1\pm s/\gamma\pm\frac{\text{v}^{2}}{2s\gamma^{3}}%
(\gamma^{2}\pm s\gamma-4m_{+}^{2}\sin^{2}\beta)\right]  ;\text{ }\sigma_{\pm
}(r)=\text{v}^{2}\left[  -\frac{2m_{\pm}}{s\gamma^{3}}\frac{\cos2\beta}%
{r}+\frac{m_{\pm}\left(  1\pm s/\gamma\right)  \sin^{2}\beta}{2s\gamma
}r\right]  .
\]
In this expression, one notes a clear dependence of the potential on the angle
$\beta,$ which is a unequivocal sign of anisotropy determined by the ubiquity
of background vector on the system.

The electric field can be obtained in a straightforward way from Eq.
(\ref{Azero4}; it looks as follows:%

\begin{align}
\overrightarrow{E}(r)  &  =-\frac{e}{2(2\pi)}\biggl\{-m_{+}\left[  \delta
_{+}-\frac{4\text{v}^{2}\cos2\beta}{s\gamma^{3}r^{2}}\right]  K_{1}%
(m_{+}r)-m_{-}\left[  \delta_{-}-\frac{4\text{v}^{2}\cos2\beta}{s\gamma
^{3}r^{2}}\right]  K_{1}(m_{-}r)+\nonumber\\
&  +m_{+}\sigma_{+}K_{0}(m_{+}r)-m_{-}\sigma_{-}K_{0}(m_{-}r)\biggr\}\overset
{\wedge}{r}.
\end{align}
Near the origin, the short-distance potential behaves like as a genuine
logarithmic function, whereas the electric field goes as a $1/r$ function:%

\begin{equation}
A_{0}(r)=-\frac{e}{(2\pi)}\left[  1+\frac{\text{v}^{2}}{2\gamma^{2}}%
(1-\cos2\beta)\right]  \ln r,\text{ \ }\overrightarrow{E}(r)=-\frac{e}{(2\pi
)}\left[  1+\frac{\text{v}^{2}}{2\gamma^{2}}(1-\cos2\beta)\right]
\frac{\overset{\wedge}{r}}{r}.
\end{equation}
Such expressions reveal that the scalar potential is always repulsive at the
origin. The presence of the anisotropy factor is not able to revert this
behavior, since v$^{2}<<s^{2}.$ Far away from the origin, the long-range
potential vanishes according to the behavior of the Bessel functions, namely:
$A_{0}(r)\rightarrow0,$ $\overrightarrow{E}(r)\rightarrow0.$ This result shows
that these solutions decay rapidly as $r\rightarrow\infty,$ revealing a strong
screening as already observed in the pure timelike case.

The graphics in Fig. 3 displays the behavior of the scalar potential compared
with the MCS-Proca one. It should be reported that the scalar potential has
shown to be always positive for all the values of parameters adopted. This
last illustration shows that the deviations from the MCS-Proca behavior are
small, a consequence of the approximation v$^{2}/s^{2}<<1,$ which does not
allow to probe the form of the scalar potential for larger values of the background.%

\begin{figure}
[h]
\begin{center}
\fbox{\includegraphics[
height=3.0364in,
width=4.0318in
]%
{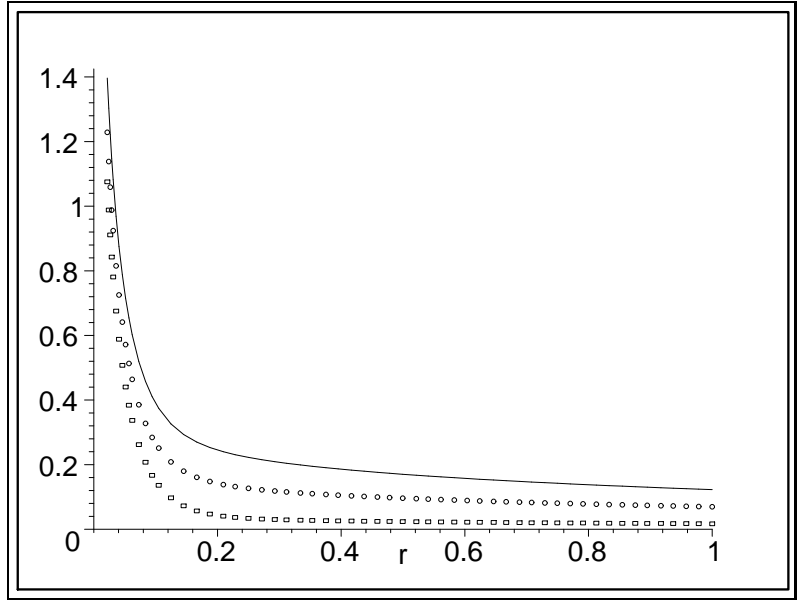}%
}\caption{Simultaneous plot of the MCS-Proca scalar potential (box dotted
curve), scalar potential for $\beta=\pi/4$ (circle dotted curve), scalar
potential for $\beta=\pi/2$ (continuos curve). The commum parameter values
are: $s=20,M_{A}=2,$v$=5.$}%
\end{center}
\end{figure}

It can be easily observed that in the absence of the background,
$\overrightarrow{\text{v}}=0,$ the scalar potential and the electric field are
reduced to the corresponding MCS-Proca ones: $A_{0}(r)=e/2(2\pi)[\left(
1+s/\gamma\right)  K_{0}(m_{+}r)+\left(  1-s/\gamma\right)  K_{0}(m_{-}r)]$,
$\overrightarrow{E}(r)=-e/2(2\pi)[\left(  1+s/\gamma\right)  m_{+}K_{0}%
(m_{+}r)+\left(  1-s/\gamma\right)  m_{-}K_{0}(m_{-}r)]\overset{\wedge}{r}$,
\ which are the solutions given in Eqs. (\ref{MCSP5}), (\ref{MCSP7}). Here,
the effect of the background vector, $\overrightarrow{\text{v}},$ appears more
clearly on the field solutions. As compared with the MCS-Proca fields ($B$ and
$\overrightarrow{E}),$ there arise supplementary terms, depending on the
background and on the angle $\beta,$ responsible for the spatial anisotropy.

As for the solution for the scalar field in the case of a purely spacelike
background, it can be obtained starting from Eq. (\ref{phi6}), which,
according to the procedure adopted so far, yields the following integral
expression:\
\begin{equation}
\varphi(r)=\frac{e}{\left(  2\pi\right)  ^{2}}\left(  \overrightarrow
{\text{v}}^{\ast}\cdot\overrightarrow{\nabla}\right)  \left[  \int_{0}%
^{\infty}\frac{\mathbf{k}d\mathbf{k}}{\left[  \mathbf{k}^{2}+R_{+}^{2}\right]
}\int_{0}^{2\pi}Qe^{ikr\cos\varphi}d\varphi-\int_{0}^{\infty}\frac
{\mathbf{k}d\mathbf{k}}{\left[  \mathbf{k}^{2}+R_{-}^{2}\right]  }\int
_{0}^{2\pi}Qe^{ikr\cos\varphi}d\varphi\right]  ,
\end{equation}
where: $Q=\left[  (s^{2}+\text{v}^{2}\sin^{2}\alpha)(s^{2}+4M_{A}^{2}%
+\text{v}^{2}\sin^{2}\alpha)\right]  ^{-1/2}.$ Making use of Eq. (\ref{App5})
and the suitable approximation, $Q\simeq1/(s\gamma)-(s^{2}+2M_{A}^{2})$%
v$^{2}\sin^{2}\alpha/(s\gamma)^{3},$ a lengthy solution may be attained for
the scalar field after a boring calculation, namely:%

\begin{equation}
\varphi(r)=-\frac{e}{\left(  2\pi\right)  }\biggl\{m_{+}\varrho_{+}%
(r)K_{1}(m_{+}r)-m_{-}\varrho_{-}(r)K_{1}(m_{-}r)-\varsigma_{+}(r)K_{0}%
(m_{+}r)+\varsigma_{-}(r)K_{0}(m_{-}r)\biggr\}\left(  \overrightarrow
{\text{v}}^{\ast}\cdot\overset{\wedge}{r}\right)  ,
\end{equation}
where:
\[
\varrho_{\pm}(r)=\left[  \frac{1}{s\gamma}-\frac{2m_{\pm}^{2}}{(s\gamma)^{3}%
}\text{v}^{2}\sin^{2}\beta\pm\frac{\text{v}^{2}}{2\left(  s\gamma\right)
^{2}}+\frac{4\text{v}^{2}\cos2\beta}{\left(  s\gamma\right)  ^{3}}\frac
{1}{r^{2}}\right]  ,\text{ }\varsigma_{\pm}(r)=\frac{m_{+}^{2}\text{v}^{2}%
}{2(s\gamma)^{2}}\left(  \pm r\sin^{2}\beta-\frac{4\cos2\beta}{\left(
s\gamma\right)  }\frac{1}{r}\right)  .
\]

Near the origin, this solution behaves as: $\varphi(r)\rightarrow
-e/(2\pi)[1/(s\gamma)^{2}]\left(  1/r\right)  \left(  \overrightarrow
{\text{v}}^{\ast}\cdot\overset{\wedge}{r}\right)  ,$ implying an attractive
character in this limit. Concerning the asymptotic behavior, the scalar
solution vanishes exponentially.

\section{Solutions for the vector potential and the magnetic field in the
static limit}

In this section, we aim at constructing the solutions for the vector potential
and magnetic field for a point-like static charge for both a timelike and
spacelike Lorentz-violating backgrounds. These solutions are achieved from the
differential equations (\ref{Maxwell4}) and (\ref{Avector}), which in the
static limit constitute a coupled system of two differential equations.

\subsection{\bigskip The external vector is purely time-like: $v^{\mu}%
=($v$_{0},0)$}

Starting from Eqs. (\ref{motion2}),(\ref{Avector}), one writes the following
system in the static limit:%

\begin{align*}
\lbrack\nabla^{2}(\nabla^{2}-s^{2}-2M_{A}^{2})+M_{A}^{4}]\overrightarrow
{A}+\text{v}_{0}\left(  \nabla^{2}-M_{A}^{2}\right)  (\nabla^{\ast}\varphi)
&  =-s\overrightarrow{\nabla}^{\ast}\rho,\\
\text{v}_{0}\nabla\times\overrightarrow{A}+\left(  \nabla^{2}-M_{A}%
^{2}\right)  \varphi &  =0,
\end{align*}
which may be decoupled in two equations for the vector potential and the
scalar field:%

\begin{align}
\lbrack\nabla^{2}(\nabla^{2}-s^{2}-2M_{A}^{2})+M_{A}^{4}]\overrightarrow{A}
&  =-s\overrightarrow{\nabla}^{\ast}\rho,\label{A2}\\
\lbrack\nabla^{2}(\nabla^{2}-s^{2}-2M_{A}^{2})+M_{A}^{4}]\left(  \nabla
^{2}-M_{A}^{2}\right)  \varphi &  =s\text{v}_{0}\nabla^{2}\rho, \label{phi2}%
\end{align}
Proposing a Fourier-transform representation for the vector potential,
$\overrightarrow{A}(r)=\frac{1}{(2\pi)^{2}}\int d^{2}\overrightarrow
{k}e^{i\overrightarrow{k}.\overrightarrow{r}}\widetilde{A}(k),$ it turns out
to be:
\begin{equation}
\overrightarrow{A}(r)=-\frac{es}{(2\pi)}C\left[  M_{+}K_{1}\left(
M_{+}r\right)  -M_{-}K_{1}\left(  M_{-}r\right)  \right]  \overset{\wedge
}{r^{\ast}}, \label{A1}%
\end{equation}
where: $C=1/\sqrt{(s^{2}-\text{v}_{0}^{2})(s^{2}-\text{v}_{0}^{2}+4M_{A}^{2}%
)},$ and the terms $M_{\pm}^{2}$ are defined in Eq. (\ref{M1}). The magnetic
field, $B=\overrightarrow{\nabla}\times\overrightarrow{A}$, stems directly
from the equation above at the form:\
\[
B(r)=-\frac{es}{(2\pi)}C\left[  M_{+}^{2}K_{0}\left(  M_{+}r\right)
-M_{-}^{2}K_{0}\left(  M_{-}r\right)  \right]  .
\]

Comparing these solutions with the MCS-Proca counterparts, one then notices
that the background does not impose any functional modification. Its role is
limited to yielding an increasing of the associated range, which can be
observed in Fig. 4.

It is simple to notice that the solutions here attained for $\mathbf{A}$ and
$B$ present the same behavior of the MCS-Proca case both near and far from the
origin. Indeed, for $r\rightarrow0$, the vector potential vanishes $\left(
\mathbf{A}\rightarrow0\right)  ,$ whereas the magnetic field behaves like a
pure logarithmic function:\
\[
B(r)\rightarrow\left(  -\frac{es}{2\pi}\right)  \ln r,
\]
in much the same way as the MCS-Proca behavior. Far from the origin, both
these fields vanish exponentially.

\bigskip In Fig. 4, it is shown a comparative illustration between the
MCS-Proca vector potential and the one given by Eq. (\ref{A1}), which
clarifies the role of the background: the larger is v$_{0}$, the larger is the
deviation from the MCS-Proca behavior (the potential becomes more positively pronounced).%

\begin{figure}
[h]
\begin{center}
\fbox{\includegraphics[
trim=0.000000in 0.000000in 0.217046in 0.225013in,
height=2.8106in,
width=3.813in
]%
{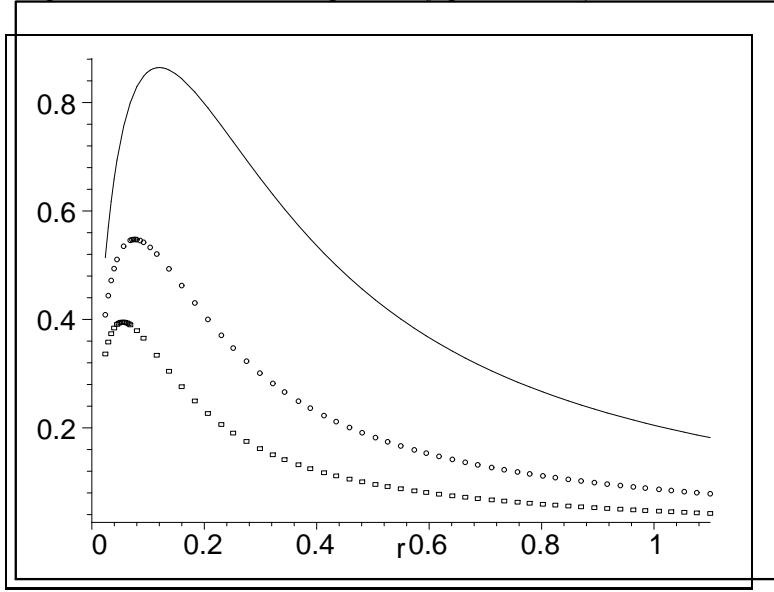}%
}\caption{Simultaneous plot for the MCS-Proca vector potential (box dotted
line), vector potential for v$_{0}=14$ (circle dotted line) and vector
potential for v$_{0}=18$ (continuos line)$,$ with $s=20,M_{A}=2.$}%
\end{center}
\end{figure}

In this section, it is still possible to derive a solution for the scalar
field in the case of a purely timelike background, which stems from Eq.
(\ref{phi2}). This solution is easily attained following the usual procedure
here adopted:
\[
\varphi(r)=\frac{e}{(2\pi)}\left(  s\text{v}_{0}\right)  \left[  B_{+}%
K_{0}\left(  M_{+}r\right)  +B_{-}K_{0}\left(  M_{-}r\right)  -(B_{+}%
+B_{-})K_{0}\left(  M_{A}r\right)  \right]  ,
\]
where the coefficients $B_{\pm}$ are given in Eqs. (\ref{b1}), (\ref{t1}).
Near the origin this solution vanishes identically, that is: $\varphi
(r)\rightarrow0.$ Far from the origin it vanishes exponentially according to
the Bessel-like asymptotic behavior.

\subsection{\ The external vector is\ purely\ space-like: $v^{\mu}%
=(0,$\textbf{v}$)$}

Starting from Eqs. (\ref{Maxwell4}),(\ref{Avector}), one attains:%

\begin{align}
\lbrack\nabla^{2}(\nabla^{2}-s^{2}-2M_{A}^{2})+M_{A}^{4}]\overrightarrow
{A}-s\overrightarrow{\nabla}^{\ast}(\overrightarrow{\text{v}}\cdot
\overrightarrow{\nabla}^{\ast}\varphi)  &  =-s\overrightarrow{\nabla}^{\ast
}\rho,\\
\lbrack(\overrightarrow{\text{v}}^{\ast}\cdot\overrightarrow{\nabla}%
)\nabla\times+M_{A}^{2}\overrightarrow{\text{v}}\cdot]\overrightarrow
{A}-s\left(  \nabla^{2}-M_{A}^{2}\right)  \varphi &  =0,
\end{align}

which can be decoupled in the two following equations:
\begin{align}
\left[  \lbrack\nabla^{2}(\nabla^{2}-s^{2}-2M_{A}^{2})+M_{A}^{4}%
]+(\overrightarrow{\text{v}}\cdot\overrightarrow{\nabla}^{\ast}%
)(\overrightarrow{\text{v}}\cdot\overrightarrow{\nabla}^{\ast})\right]
\overrightarrow{A}  &  =-s\overrightarrow{\nabla}^{\ast}\rho,\label{Avector3}%
\\
\left[  \lbrack\nabla^{2}(\nabla^{2}-s^{2}-2M_{A}^{2})+M_{A}^{4}%
+(\overrightarrow{\text{v}}\cdot\overrightarrow{\nabla}^{\ast}%
)(\overrightarrow{\text{v}}\cdot\overrightarrow{\nabla}^{\ast})]\left(
\nabla^{2}-M_{A}^{2}\right)  \right]  \varphi &  =[(\overrightarrow{\text{v}%
}^{\ast}\cdot\overrightarrow{\nabla})\nabla\times+M_{A}^{2}\overrightarrow
{\text{v}}\cdot]\nabla^{\ast}\rho. \label{phi3}%
\end{align}

The solution of Eq. (\ref{Avector3}) is given by the following integral expression:%

\begin{equation}
\overrightarrow{A}(r)=-\frac{es}{\left(  2\pi\right)  ^{2}}\overrightarrow
{\nabla}^{\ast}\left[  \int_{0}^{\infty}\frac{\mathbf{k}d\mathbf{k}}{\left[
\mathbf{k}^{2}+R_{+}^{2}\right]  }\int_{0}^{2\pi}De^{ikr\cos\varphi}%
d\varphi-\int_{0}^{\infty}\frac{\mathbf{k}d\mathbf{k}}{\left[  \mathbf{k}%
^{2}+R_{-}^{2}\right]  }\int_{0}^{2\pi}De^{ikr\cos\varphi}d\varphi\right]  ,
\end{equation}
where: $D=1/\sqrt{(s^{2}+\text{v}^{2}\sin^{2}\alpha)(s^{2}+\text{v}^{2}%
\sin^{2}\alpha+4M_{A}^{2})},$ and the factors $R_{\pm}^{2}$ are given by Eq.
(\ref{R}). \ In order to solve the integrations involved in this expression,
one should use the approximation\thinspace(\ref{App5}) supplemented by the
following one:
\[
D\simeq-\frac{1}{s\gamma}+\frac{\left(  s^{2}+2M_{A}^{2}\right)  }{\left(
s\gamma\right)  ^{3}}\text{v}^{2}\sin^{2}\alpha,
\]
Considering all that, we achieve again a lengthy result:%

\begin{equation}
\overrightarrow{A}(r)=-\frac{es}{\left(  2\pi\right)  }\biggl\{\chi
_{+}(r)K_{0}(m_{+}r)+\chi_{-}(r)K_{0}(m_{-}r)+\omega_{+}(r)K_{1}%
(m_{+}r)+\omega_{-}(r)K_{1}(m_{-}r)\biggr\}\overset{\wedge}{r^{\ast}},
\end{equation}
where:
\begin{align}
\chi_{\pm}(r)  &  =-\frac{m_{\pm}\text{v}^{2}}{\left(  s\gamma\right)  ^{2}%
}\left(  \mp\frac{2m_{\pm}}{\left(  s\gamma\right)  }\frac{\cos2\beta}%
{r}+\frac{m_{\pm}\sin^{2}\beta}{2}r\right)  ,\\
\omega_{\pm}(r)  &  =\mp m_{\pm}\left(  -\frac{1}{s\gamma}+\frac{\text{v}^{2}%
}{\left(  s\gamma\right)  ^{3}}\left(  2m_{\pm}^{2}\sin^{2}\beta\mp
\frac{s\gamma}{2}\right)  \right)  \pm\frac{4\text{v}^{2}m_{+}}{\left(
s\gamma\right)  ^{3}}\frac{\cos2\beta}{r^{2}},
\end{align}
where the spatial anisotropy determined by the fixed background becomes
manifest. Considering the behavior of the $K_{0},K_{1}-$functions near the
origin $[K_{0}(sr)\rightarrow-\ln r-\gamma_{Euler}-\ln(s/2),K_{1}%
(sr)\rightarrow1/(sr)+sr(\ln r/2+\ln(s/2)/2+(1-2\gamma_{Euler})/4]$, it is
possible to show (after some algebraic calculations) that the vector potential
vanishes in this limit $(\overrightarrow{A}(r)\rightarrow0$ for $r\rightarrow
0).$ Far away from the origin, all the terms can be neglected, so that the
vector potential also vanishes asymptotically. It is interesting to remark
that the vector potential vanishes near and far from the origin for both time-
and space-like backgrounds, recovering the pure MCS-Proca behavior. This fact
demonstrates that the background does not impose physical changes into this
potential in these two limits.

In Fig. 5, one illustrates the behavior of the vector potential compared to
the MCS-Proca vector potential. One the observes that the deviations from the
MCS-Proca behavior are very small as a consequence of the approximation
adopted, (v$/s)^{2}<<1.$ In this case, it is notorious that the background is
unable to bring about expressive modifications even at the intermediary radial
region. A similar conclusion is enclosed in Fig. 3, which exhibits the
behavior of the scalar potential (derived under the same approximation).
\begin{figure}
[h]
\begin{center}
\fbox{\includegraphics[
height=3.0364in,
width=4.0318in
]%
{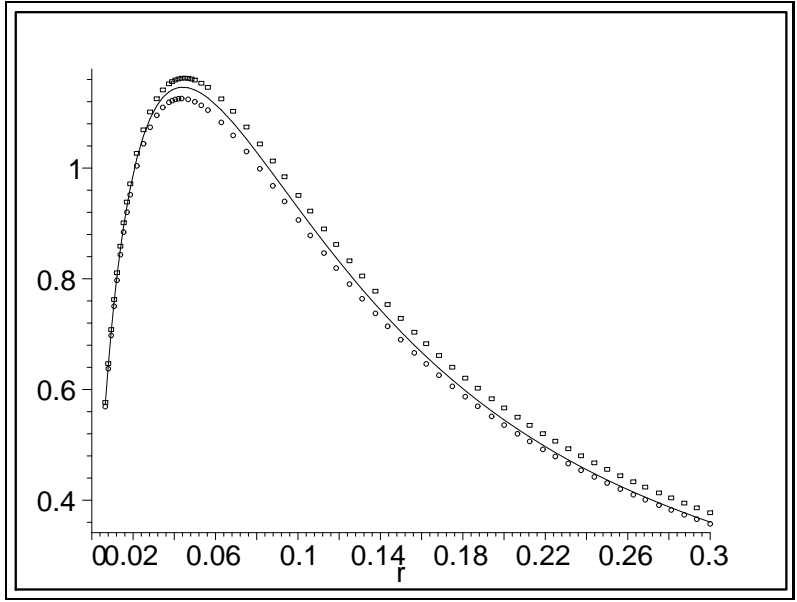}%
}\caption{Simultaneous plot for the MCS-Proca vector potential (box dotted
line), vector potential for $\beta=\pi/3$ (circle dotted line) and vector
potential for $\beta=\pi$ (continuos line)$,$ with $s=24,M_{A}=4,$v$=8.$}%
\end{center}
\end{figure}

We can now finish BY evaluating the magnetic field associated with this vector
potential, which takes the form below:%

\begin{equation}
B(r)=\frac{es}{\left(  2\pi\right)  }\biggl\{\eta_{+}(r)K_{0}(m_{+}r)+\eta
_{-}(r)K_{0}(m_{-}r)+\xi_{+}(r)K_{1}(m_{+}r)+\xi_{-}(r)K_{1}(m_{-}r)\biggr\},
\end{equation}
where:
\begin{align*}
\eta_{\pm}(r)  &  =-\frac{m_{\pm}^{2}}{s\gamma}+\frac{2m_{\pm}^{4}\text{v}%
^{2}}{\left(  s\gamma\right)  ^{3}}\sin^{2}\beta+\frac{m_{\pm}^{2}\text{v}%
^{2}}{2\left(  s\gamma\right)  ^{2}}(\pm1-2\sin^{2}\beta)\pm\frac{4m_{\pm}%
^{2}\text{v}^{2}}{\left(  s\gamma\right)  ^{3}}\frac{\cos2\beta}{r^{2}},\\
\xi_{\pm}(r)  &  =\frac{m_{\pm}^{3}\text{v}^{2}\sin^{2}\beta}{2\left(
s\gamma\right)  ^{2}}r\mp\frac{8m_{\pm}\text{v}^{2}}{\left(  s\gamma\right)
^{3}}\frac{\cos2\beta}{r^{3}}\mp\frac{2m_{\pm}^{2}\text{v}^{2}}{\left(
s\gamma\right)  ^{3}}\frac{\cos2\beta}{r}.
\end{align*}
Near the origin, this magnetic field recovers the MCS-Proca behavior
$[B(r)\rightarrow\ln r$ for $r\rightarrow0]$, while it exponentially vanishes asymptotically.

\section{Final Remarks}

Starting from a dimensionally reduced gauge invariant, but Lorentz and
CPT-violating planar model \cite{Reduction2}, derived from the
Carroll-Field-Jackiw Abelian-Higgs model \cite{Belich1}, we have studied the
extended Maxwell equations (and the corresponding wave equations for the
field-strengths and potentials) stemming from this planar Lagrangian. While
the field-strengths satisfy second-order inhomogeneous wave equations, the
potential components ($A_{0},\overrightarrow{A})$ fulfill fourth-order wave
equations, in a clear similarity to the usual behavior inherent to the pure
MCS-Proca electrodynamics. As expected, this structural resemblance is also
manifest in the solutions to these equations.

In the case of a purely timelike background, one has attained solutions for
the potentials\thinspace($A_{0},\overrightarrow{A})$ and fields$\left(
B,\overrightarrow{E}\right)  $ that behave very similarly in some respects to
the MCS-Proca counterparts. Specifically, the solutions worked out possess an
identical behavior to the MCS-Proca counterpart near and far from the origin,
revealing that the background does not affect the MCS-Proca solutions in both
these limits. The qualitative differences induced by the background appear AT
an intermediary radial region,\ in which the solutions deviate from the
MCS-Proca counterpart in a pronounced way in the case of a small Proca mass
$\left(  M_{A}/s<<1\right)  $\ or a large background $\left(  \text{v}%
_{0}\lesssim s\right)  .$\ Another effect of the background is the increasing
of the range of the solutions, also manifest in a more notorious way for
$\left(  \text{v}_{0}\lesssim s\right)  .$ Once the purely timelike
backgrounds allow the attainment of exact algebraic solutions, the value of
v$_{0}$\ may be taken as close to the value of $s$\ as possible, which points
out the role of the background. The graphs in Fig. 1 and Fig.4 illustrate
these conclusions. The MCS-Proca solutions are readily recovered whenever the
background is supposed to vanish or is very small in comparison with the other
mass parameters. A solution for the scalar field $\left(  \varphi\right)  $
was also derived, exhibiting a similar structure to the scalar potential for
both timelike and spacelike backgrounds.

In the pure space-like case, the solutions may not be obtained exactly. In
order to solve the angular integrations involved, some approximations were
considered. In general, one regards the regime in which the Chern-Simons mass
parameter is much larger than the background modulus ($s^{2}>>$v$^{2})$, so
that the solutions derived are valid to the first order in v$^{2}/s^{2}$. Due
to this approximation, there appear complex combinations of Bessel $K_{0}$ and
$K_{1}$ functions as solutions for the potentials and field strengths. All
\ these expressions depend on the angle $\beta,$ which represents the
dependence of the solutions on the 2-direction fixed by the background
($\overrightarrow{\text{v}})$. It must be stated that all these non-trivial
solutions are reduced to the simple MCS-Proca solutions in case of a vanishing
background. Analogously to the purely timelike case, the effect of the
background on the scalar potential and vector potential disappears in the
limits $r\rightarrow0,r\rightarrow\infty,$\ in which they recover the
correlate MCS-Proca behavior. It should be remarked that even at intermediary
regions these solutions exhibit only small deviations from the MCS-Proca
solutions, a direct consequence of the approximation adopted ($s^{2}>>$%
v$^{2}),$ which prevents the investigation in a situation of a large
background $\left(  \text{v}_{0}\lesssim s\right)  .$ In order to properly
analyze the solutions in this latter limit, it would be necessary to obtain
exact algebraic solutions, valid for any value of the modulus of the
background. In this case, one believes that the resulting solutions would
exhibit remarkable deviations from the MCS-Proca counterparts, as observed in
the purely timelike case. Concerning the scalar field, a solution was attained
in much the same way as done for the potentials, exhibiting a similarly
anisotropic structure. Finally, it should be pointed out that all solutions
obtained result entirely shielded, a consequence of the Proca mass, which
prevents the appearance of unescreened solutions (logarithmic ones), as it
occurs in the case of the analogous MCS-Lorentz-violating planar model solved
in Ref. \cite{Classical}.

The solutions for the scalar potential of this work put in evidence the
possibility of attaining an attractive behavior and the possible formation of
bound states. Concerning an electron-electron interaction, this issue may be
properly investigated by means of the interaction potential stemming from the
evaluation of the M\"{o}ller scattering amplitude. Such a calculation was
already carried out in the context of Lorentz-violating theories in three and
four dimensions \cite{Manojr3}. In the case of the planar model of ref.
\cite{Reduction1}, the M\"{o}ller interaction potential obtained presents an
asymptotic logarithmic behavior which represents an unreal physical
interaction in a planar dimension. In the case of the present Abelian Higgs
Lorentz-violating model, it is expected that such an evaluation yield a
screened solution, suitable for describing a real interaction in condensed
matter planar systems. This issue is a now under investigation.

\end{document}